# Quantum cascade of new correlated phases

# in trigonally warped bilayer graphene

Anna M. Seiler[1,2], Fabian R. Geisenhof[2], Felix Winterer[2], Kenji Watanabe[3], Takashi Taniguchi[4], Tianyi Xu[5], Fan Zhang[5*], R. Thomas Weitz[1,2,6*]

[1] 1st Physical Institute, Faculty of Physics, University of Göttingen, Friedrich-Hund-Platz 1, Göttingen 37077, Germany

[2] Physics of Nanosystems, Department of Physics, Ludwig-Maximilians-Universität München, Geschwister-Scholl-Platz 1, Munich 80539, Germany

[3] Research Center for Functional Materials, National Institute for Materials Science, 1-1 Namiki, Tsukuba 305-0044, Japan

[4] International Center for Materials Nanoarchitectonics, National Institute for Materials Science, Tsukuba, Japan

[5] Department of Physics, University of Texas at Dallas, Richardson, TX, 75080, USA

[6] Center for Nanoscience (CeNS), Schellingstrasse 4, Munich 80799, Germany

*Corresponding author. Email: thomas.weitz@uni-goettingen.de, zhang@utdallas.edu

**Summary Paragraph:**

Divergent density of states offers the unique opportunity to explore a wide variety of correlated electron physics. In the thinnest limit, this has been predicted and verified in the ultra-flat bands of magic-angle twisted bilayer graphene[1–5], the band touching points of few-layer rhombohedral graphite[6], and the lightly doped rhombohedral trilayer graphene[7]. The simpler and seemingly better understood Bernal bilayer graphene is also susceptible to orbital magnetism-driven phases at charge neutrality[6], such as layer antiferromagnet[8] and quantum anomalous Hall octet[9]. Here we report the discovery of a cascade of novel correlated phases in the vicinity of electric-field-controlled Lifshitz transitions[10,11] and van Hove singularities[12] in trigonally warped bilayer graphene. We provide compelling evidence for the observation of Stoner ferromagnets – half and quarter metals[13,14]. More prominently, we identify signatures consistent with a topologically nontrivial Wigner-Hall crystal[15] at zero magnetic field and its transition to a trivial Wigner crystal, as well as two correlated metals whose behavior deviates from standard Fermi liquids. Our results in this reproducible, tunable, simple system opens a new chapter for studying strongly correlated electrons.

**Main text:**

Electron-electron interactions can give rise to macroscopic quantum phenomena such as magnetism, superconductivity, and quantum Hall (QH) effect. Well known is that the interaction effects can be effectively and efficiently boosted near the energy where electrons' density of states (DOS) diverges. One remarkable example is twisted bilayer graphene, in which the Dirac minibands become extremely flat when the twist angle is near one degree – the so-called magic angle[5]. Indeed, orbital ferromagnetism[16], quantum anomalous Hall (QAH) effect[17], and nematic superconductivity[18] have been observed in this delicately designed system. In fact, strongly interacting behavior has also been discovered in the simpler, naturally occurring, Bernal bilayer graphene (BLG), particularly near charge neutrality. Under a high magnetic field, its bands quantize into Landau levels (LL) that are exactly flat, and both fractional QH effect and QH ferromagnetism have been reported[9,19,20]. Near zero magnetic field, its quadratic band touching points that each enjoy a nontrivial winding number of $\pm 2$ are susceptible to spontaneous gaps driven by topological orbital magnetism[6], as evidence by a QAH octet observed in free-standing BLG[9]. Interestingly, when lightly doped the Fermi surface of BLG (and its rhombohedral cousins) becomes warped, attributed to the interlayer next-nearest-neighbor coupling[21]. The winding number and the $C_{3z}$ rotational symmetry dictate a deformation of each quadratic Dirac cone into one central cone and three surrounding ones. An electric field can gap these cones and even flip the central one. These produce multiple Lifshitz transitions and a van Hove singularity tunable by the electric field and the charge density. However, to date correlated phases mediated by this trigonal warping effect have escaped experimental observation in BLG, though ferromagnetism and superconductivity were reported in lightly doped rhombohedral trilayer graphene (RTG) very recently[13,14].

Here we show that, at large electric fields, gate tunable correlated insulating and metallic phases emerge in lightly doped high-quality Bernal BLG, without the presence of a moiré potential. The two novel insulating phases are consistent with a Wigner crystal and an unprecedented Wigner-Hall crystal, respectively, and the novel metallic phases are most likely fractional metals that develop LL and correlated metals that show no signs of Landau quantization.

The investigated BLG flakes are encapsulated in hexagonal Boron Nitride (hBN) and equipped with graphite top and bottom gates as well as graphite contacts in a two-terminal configuration (Supplementary Fig. 1 and Methods). By varying both gate voltages, the charge carrier density $n$ and the perpendicular electric field $E$ can be tuned. Figure 1a shows the two-terminal conductance $G$ measured at zero magnetic field and a temperature $T$ of 10 mK. Due to the screening of Coulomb interaction by hBN, the interaction induced spontaneously gapped phase at $E = 0$ and $n = 0$ is absent, unlike in freestanding BLG[9,22,23]. Whereas the horizontal line of decreased conductance crossing $E = 0$ stems from our graphite contacts (Methods). The most striking physics is observed at high electric fields outside this region (see below). Due to the magnetic field and electric field dependent contact resistance, the measured conductivities are given in arbitrary units (a.u.), and nontrivial topological phases are identified by their slopes in a fan diagram.

**Magnetotransport in bilayer graphene**

To establish our sample system, we first discuss magnetotransport features observed previously[10,11,22,24,25]. At charge neutrality, an electric-field induced band gap results in the well-known conductance minimum[22]. A crystallographic alignment between the BLG, hBN, and graphite contacts can be excluded, given the clear absence of side conductance minima at higher density[26,27]. While this electric field gap seems symmetric in density for low electric fields, an electron-hole asymmetry increases with the electric field[28], attributed to the remoter interlayer hoppings beyond the next-nearest-neighbor ones (whereas the latter are responsible for the trigonal warping)[29]. We focus our discussion on the hole side where the most distinct features are observed. An onset of similar physics however can also be observed on the electron side (Methods, Supplementary Fig. 2). Furthermore, the conductance map displays slight asymmetry in electric field, likely due to the different cleanliness levels of the two sides of the bilayer. Nevertheless, the main signatures described hereafter are the same for both the negative and positive electric fields.

At zero electric field we observe the full splitting of the lowest LL octet as a function of the perpendicular magnetic field $B$ expected for clean samples (Fig. 1b, Supplementary Fig. 3a) together with electric field induced LL crossings (inset Fig. 1b and Supplementary Fig. 3b,c). Notably, QH states start to emerge at very low magnetic fields below 0.2 T, demonstrating the high quality of our device (Supplementary Fig. 3a)[22,24,25]. At a large electric field the trigonal warping effect becomes more visible since the central hole pocket is compressed into an electron pocket at a lower energy (Fig. 1d,e). At very low doping the Fermi surface is consequently composed of three disconnected hole pockets that move apart with increasing the electric field. At slightly higher doping these three pockets connect, and the Fermi surface undergoes the first Lifshitz transition and then becomes an annulus, i.e., an inner electron pocket and an outer hole pocket. As the doping increases further, the hole pocket enlarges whereas the electron pocket contracts and disappears in the second Lifshitz transition[11,12,30]. Figure. 1c shows a conductance map as a function of $n$ and $B$ at $E$ = 0.6 V/nm. At very low doping the three disconnected hole pockets result in a three-fold LL degeneracy per spin-valley, and only the $v$ = -3 and -6 QH ferromagnetic states emerge at magnetic fields below approximately 1.5 T (Fig. 1c, Supplementary Fig. 4), whereas above ~ 2.5 T all the integer QH ferromagnetic states are resolved. In between 1.5 T and 2.5 T there are two crossings between the $v$ = -4 and -5 QH states, originating from the emergence and contraction of the inner electron pocket whose LLs do not mix with the hole LLs[10,11] (Supplementary Fig. 5). The corresponding densities $n$ = - 2.1 x $10^{11}$ cm$^{-2}$ and $n$ = -2.7 x $10^{11}$ cm$^{-2}$ at $E$ = 0.6 V/nm mark the locations of the two Lifshitz transitions where the DOS diverges and jumps, respectively (Fig. 1c, Supplementary Fig. 4). Such a saddle-point van Hove singularity is known to be a fertile ground for interaction driven phases even in the absence of magnetic field. Indeed, at magnetic fields below 1.3 T the conductivity pattern strongly deviates from any conventional LL sequence.

For example, in sharp contrast to the established single-particle picture[22,31], a complex non-monotonous electric field and density dependent variation of the conductance emerges for $E$ > 0.1 V/nm even at $B$ = 0 (Fig. 1a). This unexpected phase diagram of Bernal BLG stems from the intricate interplay between its electron-electron interaction and trigonal warping[10,12], and the identification of its nature and topology is the main result of this work.

**Stoner spin-valley ferromagnetism**

The prominent steps in the conductance that already start to appear at low electric fields above 0.1 V/nm in Fig. 1a are highlighted in Fig. 2a in a zoom of the conductance map for positive electric fields and hole doping. For further discussions we label regions of constant conductance with letters A - E. Their electric field and density dependence is reminiscent of those phases that were observed recently in RTG and assigned to half and quarter metallic phases with spontaneous spin-valley polarizations[13], analogous to the textbook Stoner ferromagnets. Prior to this, magnetic phases have been proposed in electric field gapped BLG[12,32], given the divergent DOS in the vicinity of the band edges. Yet, such magnetic phases have escaped previous identification. Indeed, the experimental signatures observed in our BLG agrees very well with exchange-interaction-driven ferromagnetic phases that can be well resolved at $B$ = 0.6 T (Fig. 2b,d).

Close to the valence band edge, the $\nu$ = -3 and -6 QH states emerge (Fig. 2b, Supplementary Fig. 4). We attribute to the three-fold degeneracy to the trigonally warped Fermi surface. This phase (phase A) is most consistent with a quarter metal with spin and valley both fully polarized, otherwise the first observed QH states should be the $\nu$ = -6 and -12 states (for a half metal) or the $\nu$ = -12 and -24 states (for the full metal). At larger densities phase B exhibiting two-fold LL degeneracy emerges (Fig. 2b,d). Its strong magnetic hysteresis (Fig. 2c) is indicative of its spin or valley polarization, and this phase is consistent with a half metal[8]. Likewise, phase E is most likely a full metal, because of its four-fold LL degeneracy (Fig. 2b,d) and the absence of magnetic hysteresis (Fig. 2c). Notably, there appears to be a secondary phase transition within phase E in which the LLs cross (Fig. 2d). Unlike phases A, B and E, phases C and D both exhibit conductance oscillations versus $E$ instead of $n$, possibly due to the partial spin-valley polarizations with annular fermi surfaces. Because of BLG's smaller DOS compared to that of RTG, phases A - E emerge in narrower density ranges. Universally, these Stoner phases extend to higher densities with increasing the electric field, since the van Hove singularity can be enhanced by increasing the electric field. Likely, there may exist extra Stoner phases in BLG, yet they are too close in density to phases A - E to be disentangled.

**New correlated metallic and insulating phases**

Besides the phases consistent with the Stoner physics, we have also identified several phases in the conductance map that compete with the Stoner phases yet have not been previously identified in graphene systems. We focus our discussions hereafter on these new phases in the large negative electric field range (see Fig. 2a for the $E > 0$ data), and there are at least four distinct phases at $B = 0$ (labeled I – IV, Fig 3a) that can be identified by steps in the conductance and stabilities in the space of $E$, $n$, and $B$. All four phases were also identified in a second device (Methods, Supplementary Fig. 6). For low doping, the conductance increases with increasing density until it reaches a plateau (phase I). The conductance then drops significantly (phase II) with further increasing density until it increases again (phase III). Additionally, another plateau of higher conductance can be observed at a still higher density (phase IV). While in general these phases are all unstable against the Stoner phases at large magnetic fields ($B > 0.5 – 1.5$ T), their mutual phase boundaries and those to the Stoner phases exhibit sharp conductance changes at small magnetic fields (Fig. 3b). To identify the nature of these novel phases, we map their detailed stabilities against $n$ as functions of $B$ (Fig. 3c-f) and $T$ (Fig. 4). To enhance the visibility of the transitions between these phases, the density derivative of the conductance map is shown as a function of $B$ and $n$ (Fig. 3e-f).

Phase I is an island of relatively high conductance at very low densities close to the valence band edge. It competes with phase A (a quarter metal) in which the single-particle Fermi surface consists of three disconnected hole pockets at each spin-valley. Its intimate connection to the trigonal warping is also evidenced by the observation that its center and extend in density increases with increasing the electric field (Fig. 3).

Surprisingly, although at low electric fields LLs are even visible down to 0.2 T (Fig. 1b), phase I shows no signs of Landau quantization but instead an electric field dependent phase boundary in the space of $n$ and $B$ against phase A (Fig. 3c-f, Supplementary Fig. 7, Supplementary Fig. 8a). For example, phase I is stable up to 0.5 T at $E$ = -0.8 V/nm but only to 0.2 T at $E$ = -0.6 V/nm. Moreover, phase I exhibits intrinsic orbital magnetism, as evidenced by its slight magnetic hysteresis (Supplementary Fig. 9). The electric field not only enhances the stability of phase I in magnetic field but also in temperature: temperature dependent measurements (Fig. 4 and Supplementary Fig. 10) reveal that phase I is stable up to 4 K at $E$ = -0.8 V/nm but only to 3 K at $E$ = -0.6 V/nm. At $E$ = -0.8 V/nm (Supplementary Fig. 11b), its resistance increases with increasing temperature above 1 K, consistent with a correlated metal, yet surprisingly it reaches a residual below 1 K. The deviation from the standard $T^2$ scaling (Supplementary Fig. 11b,c) is suggestive that phase I is likely a non-Fermi liquid, which requires further examination. A similar absence of temperature dependence appears to also be observed above a superconducting phase transition in RTG[8].

At slightly higher densities an abrupt transition (Fig. 4a,b) from phase I to phase II of lower conductance is visible. Remarkably, phase II shows an increasing resistance with decreasing temperature (Fig. 4 and Supplementary Fig. 10) indicative of an energy gap, but its low-temperature conductance matches that of the $\nu$ = -2 QH state at $B$ = 0 with a slope of -2 in the fan diagram (Fig. 3c-f) and a prominent hysteresis in magnetic field (Supplementary Fig. 9). These together provide strong evidence for a QAH phase with a Chern number of 2. In sharp contrast to the QAH octet recently reported in freestanding BLG[9], this QAH phase II is stable at finite densities that increase with increasing the electric field (Fig. 3a,b and Supplementary Fig. 12), which implies an intimate connection to the trigonal warping (likely to the van Hove singularity as implied in Supplementary Fig. 4). While it can be suppressed by the magnetic field (Fig. 3d and f, Supplementary Fig. 8b) and temperature (Fig. 4), this QAH phase II is present at $B$ = 0 T for 0.3 V/nm < $E$ < 0.8 V/nm (Supplementary Fig. 7), While below ~0.6V/nm a larger electric field stabilizes it against magnetic field, for $E$ > 0.8 V/nm phase II becomes less stable to the magnetic field (Supplementary Fig. 7,8) indicative of its partial layer polarization.

The observation of a QAH phase emerging at $B$ = 0 but $n \neq 0$ is extraordinary. This indicates an energy gap opening at densities where the non-interacting phase and even the mean-field Stoner phases would not be gapped. Since the density range is extremely low and incommensurate with any fractional filling of BLG bands, Mott insulator phases present in graphene moiré systems can be excluded first. Moreover, incommensurate density wave (IDW) phases prone to Fermi surface nesting and/or coupling to atomic structure can also be excluded for three reasons. (i) Fermi surface nesting is absent, and any structure distortion is unlikely. (ii) The metallic nature of IDW phases is against the observed insulating behavior here, not to mention their usual trivial Chern numbers. (iii) Whereas IDW phases are incompressible, phase II is as compressible as the observed Stoner phases. Interestingly, phase II is consistent with a Wigner-Hall crystal phase, i.e., an electron crystal with a quantized Hall conductance. Such a Wigner-Hall crystal at a finite magnetic field was originally proposed by Tešanović, Axel, and Halperin[15], and here phase II may be viewed as the anomalous counterpart at $B$ = 0.

This Wigner-Hall physics can be captured the Diophantine equation $n = \nu n_\phi + \eta A_0^{-1}$, where $\nu$ is the total Chern number, $n_\phi = eBh^{-1}$ is the density of magnetic flux quanta, $\eta$ is the band filling, and $A_0$ is the unit cell area of electron crystal[15]. A Wigner crystal has $\nu = 0$ and $\eta \neq 0$, while a Hall crystal has $\nu \neq 0$ and $\eta = 0$. The more unusual case for $v \neq 0$ and $\eta \neq 0$ is the Wigner-Hall crystal. For phase II, one possible scenario is $= -2$ and $\eta = 2$: the doped holes spontaneously crystalize into a honeycomb lattice realizing the spin degenerate Haldane model[6,33], or they form a Wigner crystal on top of the undoped system that is in one state of the QAH octet[6,9].

Phase II is unstable to another gapped phase III at slightly higher densities. The two phases compete in nearly the same density range (Fig. 3a-d), and phase III dominates for $E > 0.8$ V/nm at $B = 0$ (Fig. 3a). Phase III displays neither a slope in the fan diagram nor a sign of any LL, and in fact it can be suppressed by the magnetic field (Fig. 3c-f). The observation of a magnetic hysteresis (Supplementary Fig. 9) hints an orbital magnetic order of this phase. Along the same line of arguments, a potential candidate phase for phase III is a Wigner crystal with $v = 0$ and $\eta \neq 0$. At larger densities the system enters into a metallic phase IV of high conductance but unstable at the magnetic fields leading to the Stoner phases. In the fan diagram (Fig. 3c-f) phase IV has a similar shape to phase I but a larger critical magnetic field that increases with increasing the electric field (Supplementary Fig. 8d). Phase IV exhibits a more pronounced magnetic hysteresis than phase I indicative of orbital magnetism (Supplementary Fig. 9). Like phase I, the resistance of phase IV decreases with decreasing temperature and deviates from the standard $T^2$ scaling (Supplementary Fig. 11). This suggests phase IV to be another correlated metal, likely a non-Fermi liquid that requires future examinations.

**Discussion and outlook**

Our results reveal a cascade of density and field dependent correlated phases in Bernal BLG. Transport evidence is provided for Stoner ferromagnets including the half metal and the quarter metal, electron crystals including the conventional Wigner crystal (topologically trivial) and the unprecedented Wigner-Hall crystal (topologically non-trivial), and two correlated metals whose behavior deviates from standard Fermi liquids. These novel phases are driven by the complex interplay between the electron-electron interactions, the trigonal warping effect, and the interlayer electric field. However, deciphering the origin of each phase remains challenging, and deeper understanding of this strongly correlated electron system calls for new experiments and more theoretical work. For example, multi-terminal measurements could test the spontaneous time-reversal symmetry breaking in general and better resolve the QAH effect in the Wigner-Hall crystal, nonlinear Hall effect[34] could determine the spontaneous rotational symmetry breaking, and ultraviolet diffraction or scanning tunneling microscopy[35] could probe the crystal structures of electron crystals. Finally, given that in RTG and twisted BLG superconductivity occurs close to the parent fractional metals[13,14] and correlated insulators[2,36], similar effects could be observed in Bernal BLG possibly when the thickness of a hBN flake serving as a dielectric is reduced.

During the submission of our work, we became aware of two experimental works by H. Zhou, et al.[37] and La Barrena et al.[38] on trigonally warped bilayer graphene. They both reported the metallic Stoner phases but not the new correlated phases with metallic and insulating behavior.

**Acknowledgements:** We thank V. Falko for fruitful discussions. R.T.W. and A.M.S. acknowledge funding from the Center for Nanoscience (CeNS) and by the Deutsche Forschungsgemeinschaft (DFG, German Research Foundation) under the SFB 1073 project B10 and under Germany's Excellence Strategy-EXC-2111-390814868 (MCQST). T.X. and F.Z. acknowledge support from the Army Research Office under grant number W911NF-18-1-0416 and the National Science Foundation under grant numbers DMR-1945351 through the CAREER programme, DMR-2105139 through the CMP programme, and DMR-1921581 through the DMREF programme. K.W. and T.T. acknowledge support from the Elemental Strategy Initiative conducted by the MEXT, Japan (Grant Number JPMXP0112101001) and JSPS KAKENHI (Grant Numbers 19H05790, 20H00354 and 21H05233).

**Author contributions:** A.M.S. fabricated the devices and conducted the measurements and data analysis. K.W. and T.T. grew the hexagonal nitride crystals. T.X. and F.Z. contributed the theoretical part. All authors discussed and interpreted the data. R.T.W. supervised the experiments and the analysis. The manuscript was prepared by A.M.S., F.Z. and R.T.W. with input from all authors.

**Corresponding authors:** Correspondence to R.T. Weitz (thomas.weitz@uni-goettingen.de) or F. Zhang (Zhang@utdallas.edu)

**Competing interests:** Authors declare no competing interests.

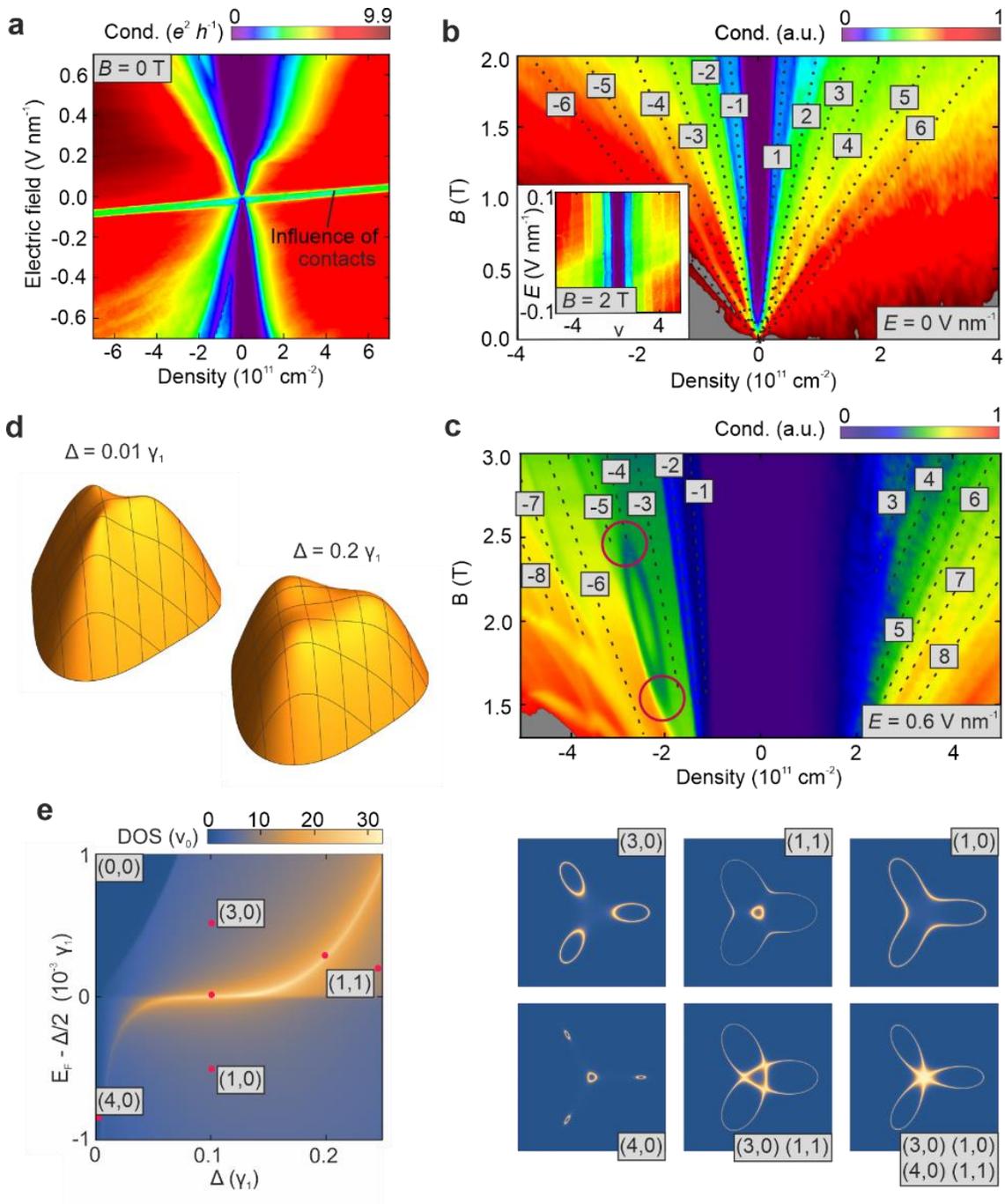

**Fig. 1. Magnetotransport in bilayer graphene. (a)** Conductance map as a function of $E$ and $n$ measured at $B$ = 0 T and a temperature of 10 mK. **(b,c)** Fan diagram of the conductance at $E$ = 0 V/nm **(b)** and $E$ = 0.6 V/nm **(c)**. Landau levels are traced by black dotted lines and corresponding filling factors ($\nu$) are indicated by Arabic numerals. Inset in (a): conductance as a function of $\nu$ and $E$ at $B$ = 2 T. In (c) the two crossings between the $\nu$ = -4 and -5 QH states are highlighted by red circles. **(d)** Calculated trigonally warped valence bands for the screened interlayer potential difference (induced by $E$) $\Delta = 0.01\gamma_1$ and $0.2\gamma_1$, respectively. **(e)** Left panel: calculated density of states (DOS) as a function of $\Delta$ and the Fermi energy $E_F$ (in units of the constant DOS $\nu_0$ of the BLG without trigonal warping). Right panels: calculated trigonally warped Fermi surfaces indicated by the red dots in the left panel. The Fermi surface topology is classified by two invariants ($i, j$), where $i$ and $j$ are the numbers of hole and electron pockets, respectively. In the left panel, the brightest line manifests the van Hove singularity at the transition between (3, 0) and (1, 0) for $\Delta < 0.1\gamma_1$ or (1, 1) for $\Delta > 0.1\gamma_1$. The $E_F = \Delta/2$ line at $\Delta > 0.1\gamma_1$ displays the (dis)appearance of the central electron pocket at the transition between (1, 1) and (1, 0).

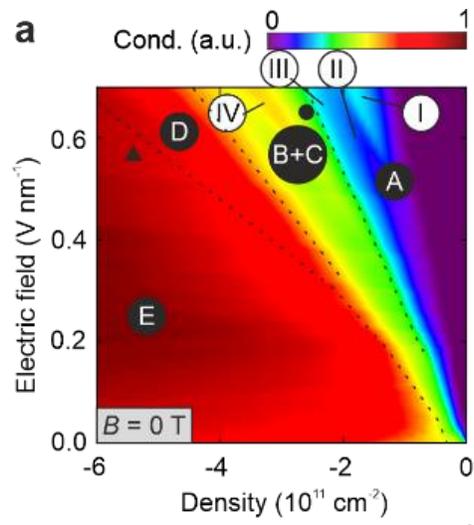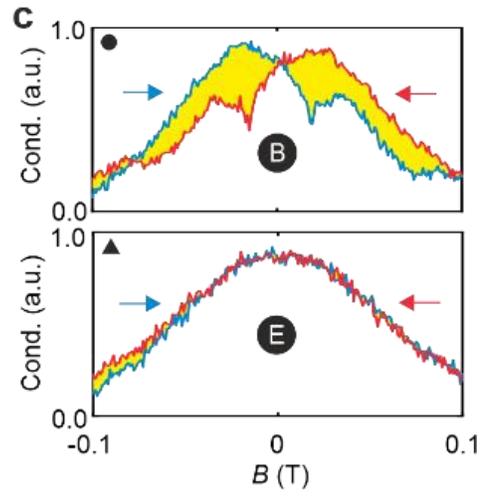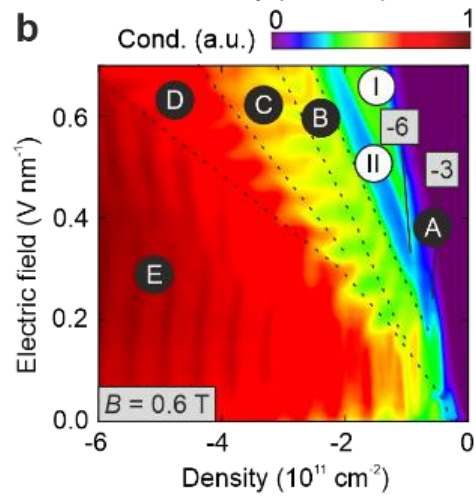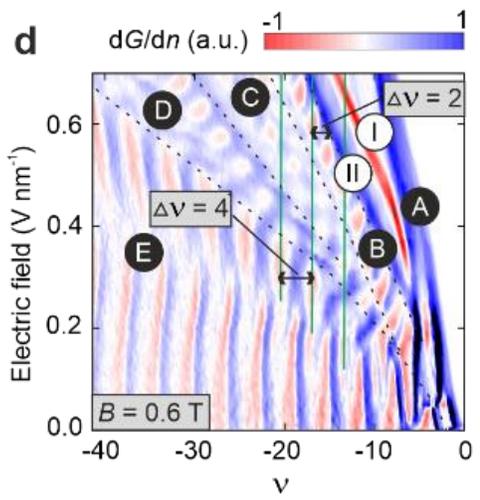

**Fig. 2. Stoner ferromagnetism in bilayer graphene. (a,b)** Conductance map as a function of $E$ and $n$ measured at $B = 0$ T **(a)** and $B = 0.6$ T **(b)** for the valence band and positive electric fields. Stoner phases are labeled with letters A - E and separated by dashed lines. New correlated phases are labeled by roman numerals I - IV. The $\nu = -3$ and -6 QH states are labeled by arabic numerals in (b). **(c)** Hystereses of the conductance as a function of $B$ in phase B ($n = -2.6 \times 10^{11}$ cm$^{-2}$, $E = 0.65$ V/nm) and in phase E ($n = -5.5 \times 10^{11}$ cm$^{-2}$, $E = 0.5$ V/nm). The forward sweeps are shown in blue while the reverse ones in red. The hysteresis loop areas are shaded in yellow for clarity. **(d)** Derivative of conductance as a function of $\nu$ and $E$. Four-fold degenerate Landau levels in phase E are indicated by green lines.

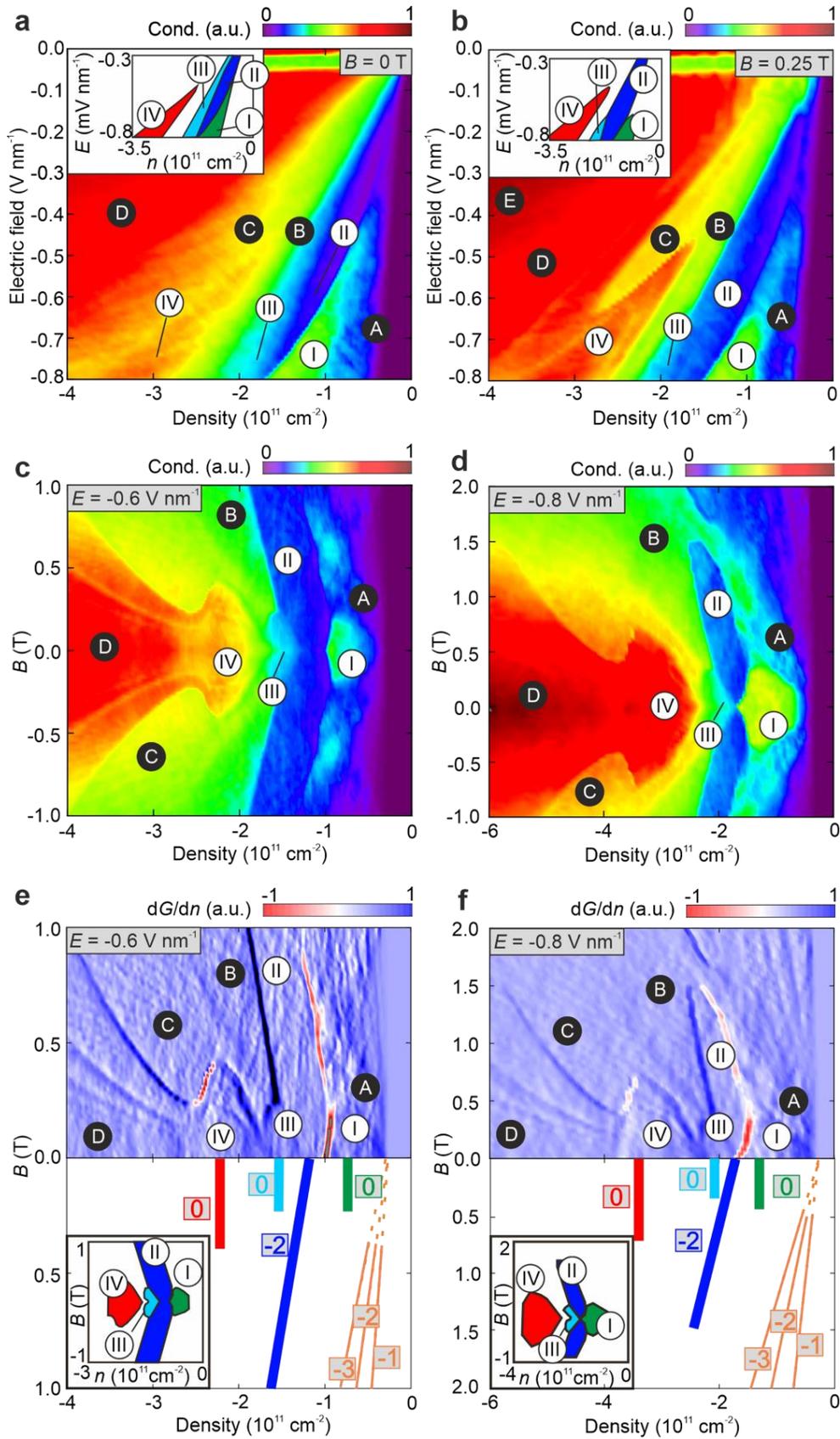

**Fig. 3. New correlated phases at high electric fields. (a, b)** Conductance map as a function of $E$ and $n$ measured at $B = 0$ T **(a)** $B = 0.25$ T **(b)** for the valence band and negative electric fields. Stoner phases are labeled with letters A - E and separated by dashed lines. New correlated phases are labeled by roman numerals I - IV. Insets: schematics showing phases I - IV. **(c-f)** Conductance map as a function of $B$ and $n$ at $E = -0.6$ V/nm **(c)** and $E = -0.8$ V/nm **(d)** and its derivative **(e, f)**. The slopes of the lowest integer QH states and of the phases I - IV are traced by lines in the schematics. The corresponding slopes are indicated by arabic numerals. The lines are solid if the states are present and dashed otherwise. Insets: schematics showing phases I - IV.

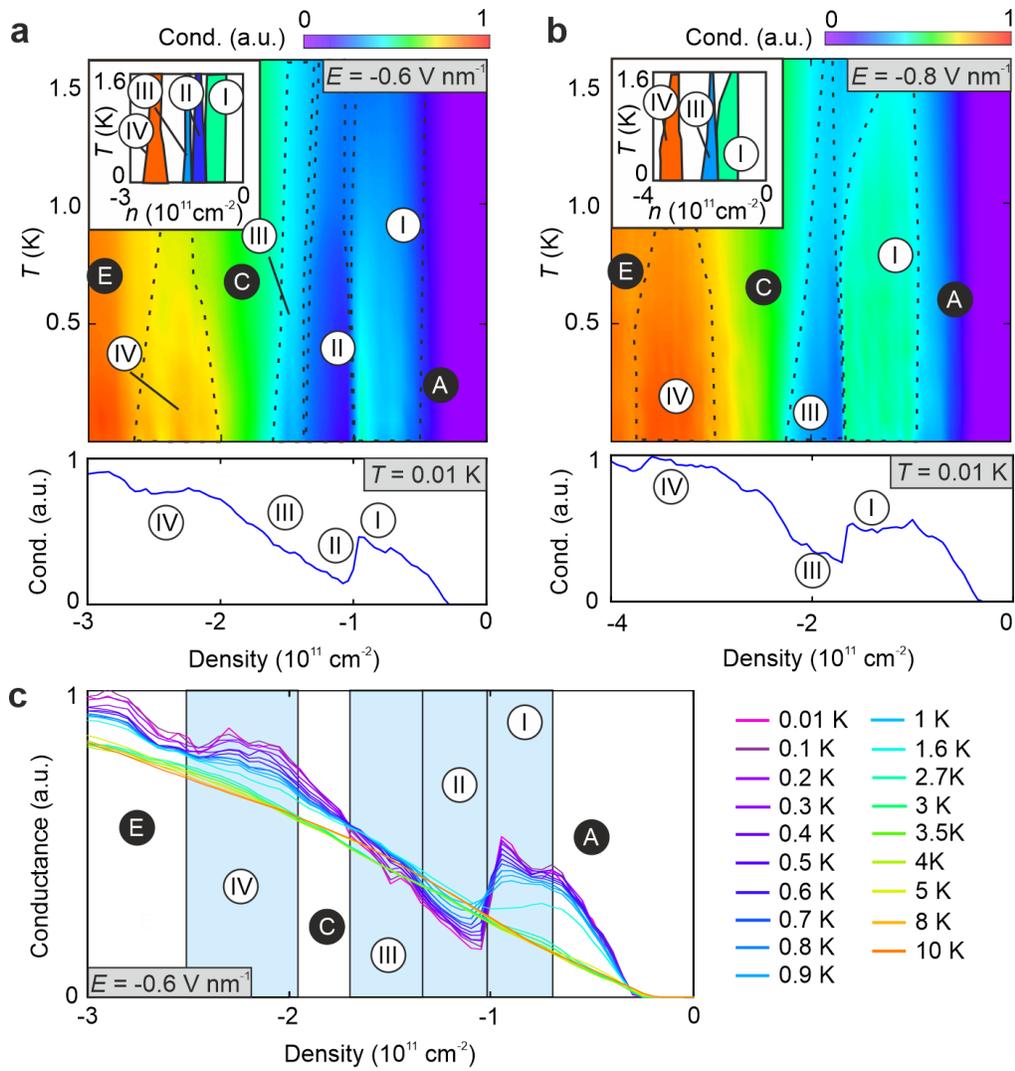

**Fig. 4. Temperature stability of correlated phases. (a, b)** Conductance at a function of *n* and temperature at *E* = -0.6 V/nm **(a)** and *E* = -0.8 V/nm **(b)** at *B* = 0 T. The new correlated phases I - IV are outlined by black dashed lines. Insets: schematics showing phases I - IV. **(c)** Conductance as a function of *n* at *E* = -0.6 V/nm and different temperatures. The new correlated phases I - IV are highlighted in light blue.

**Methods**

**Device fabrication**

Bilayer graphene flakes, graphite flakes, and hBN flakes, synthesized as described previously[39], were exfoliated on Si/SiO$_2$ substrates and subsequently identified with optical microscopy. Raman spectroscopy was used to confirm the number of layers of the bilayer graphene flakes. The encapsulated bilayer graphene devices were fabricated using the same dry transfer method as described in Ref.[40]. An hBN flake, two few-layer graphite flakes serving as contacts, a bilayer graphene flake, a lower hBN flake and a graphite flake serving as a bottom gate were picked up and then placed onto a highly doped Si/SiO$_2$ substrate. Afterwards, the samples were annealed in a vacuum chamber for 200 °C for 12 h. In a second step, a top hBN flake and, subsequently, a graphite flake serving as a top gate were picked up and released onto the annealed heterostructures. The thicknesses of the hBN flakes serving as dielectrics were determined to be 34 nm (top dielectric of device A), 42 nm (bottom dielectric of device A), 13 nm (top dielectric of device B), and 32 nm (bottom dielectric of device B) by using atomic force microscopy. Metal contacts (Cr/Au, 5 nm /60 nm for device A and 2 nm/ 45 nm for device B) that connect the graphite contacts and gates with larger pads were then structured using electron-beam lithography and were evaporated onto the sample. Optical images as well as a schematic of our devices is shown in the Supplementary Fig. 1.

**Electrical measurements**

All quantum transport measurements were conducted in a dilution refrigerator equipped with a superconducting magnet. Unless stated otherwise the sample temperature was at 10 - 20 mK. Two-terminal conductance measurements were performed using an AC bias current of 1-10 nA at 78 Hz using Stanford Research Systems SR865A and SR830 lock-in amplifiers. Home built low-pass filters were used to reduce high frequency noises. Gate voltages were applied using Keithley 2450 SourceMeters. The density and electric field were varied by varying the top and bottom gate voltages[9]. Using graphite contacts allows us to avoid etching into the insulating hBN layers as usually required for metal contacting, which would lower the accessible electric field range. However, using graphite contacts makes it technically demanding to use four-probe contacts, and two-point measurements were thus used here. The horizontal line of decreased conductance crossing $E = 0$ (Fig. 1a) stems from the part of the BLG flake that is located below the graphite contacts. As the graphite contacts are top contacts they screen the field of the top gate and the resistance in this region only depends on the applied bottom gate voltage.

**Temperature dependent measurements**

The temperature range between 1.6 K and 2.7 K is not accessible in our dilution refrigerator. Thus, we only show temperature dependent data up to 1.6 K in Fig. 4a,b. Noninterpolated temperature dependent data for temperatures up to 5 K are shown in Supplementary Fig. 10.

**Device quality**

Supplementary Fig. 3a shows the conductance plotted as a function of the density and the magnetic field at $E = 0.05$ V/nm. It can be clearly seen that the lowest QH states start to emerge at very low magnetic fields of 0.2 T, demonstrating the high quality of our device[9]. At higher magnetic fields all integer filling factors appear due to spontaneous symmetry breaking[22]. Supplementary Fig. 3b shows the conductance as a function of the density and the electric field at $B = 2$ T. All integer QH states are clearly visible. In addition, one can see several transitions in the electric field that mark the collapse of different LLs due to the interplay of spin and valley splitting[24]. We further investigated the $\nu = 0$ QH state as a function of the electric field and the magnetic field (see Supplementary Fig. 3c). We observed unusual sharp conductance peaks marking the transition between the canted antiferromagnetic (CAF) phase and the fully layer polarized (FLP) phase. This underlines the high quality of our device[41].

**Additional magnetotransport data**

In the main text we have focused on the hole side where most distinct features were observed. While we did not find signatures of phases I – IV on the electron side (see Fig. 1a), there could be Stoner phases in the conduction band. In Supplementary Fig. 2 we show the derivative of the conductance as a function of the filling factor $v$ ($v$>0) and the electric field and highlight regions with two-fold and four-fold degeneracies that potentially correspond to half and full metal phases. For completeness, the conductance as well as its derivative as functions of the charge carrier density and the magnetic field are shown in Supplementary Fig. 7 for different electric fields not shown in the main text. Furthermore, the derivative of the conductance as a function of the charge carrier density and the magnetic field at an electric field of -0.8 V/nm is shown for a second device (Device B) in Supplementary Fig. 6. Even though Device B is not as clean as Device A we still identified all phases discussed in the main text. The four phases I - IV show approximately the same density, electric field, and magnetic field behavior in both devices (Supplementary Fig. 8).

**Data availability**

The data that support the findings of this study are available from the corresponding authors upon reasonable request.

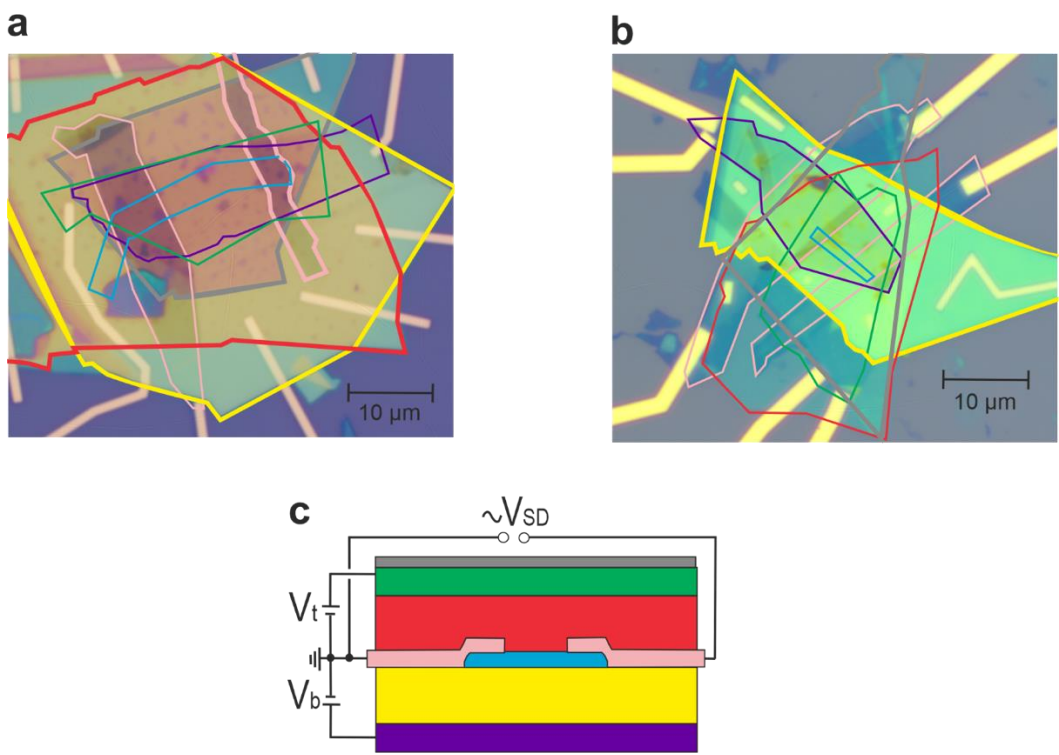

**Supplementary Fig. 1. Bilayer graphene devices studied in this work. (a, b)** Optical images of the device A (a) and device B (b)**.** The top hBN is encircled in grey, the upper graphite flake in green, the upper hBN flake in red, the graphite contacts in pink, the bilayer graphene flake in blue, the lower hBN flake in yellow and the lower graphite flake in purple. **(c)** Schematic of the bilayer graphene devices. The colors of different flakes match those in (a) and (b).

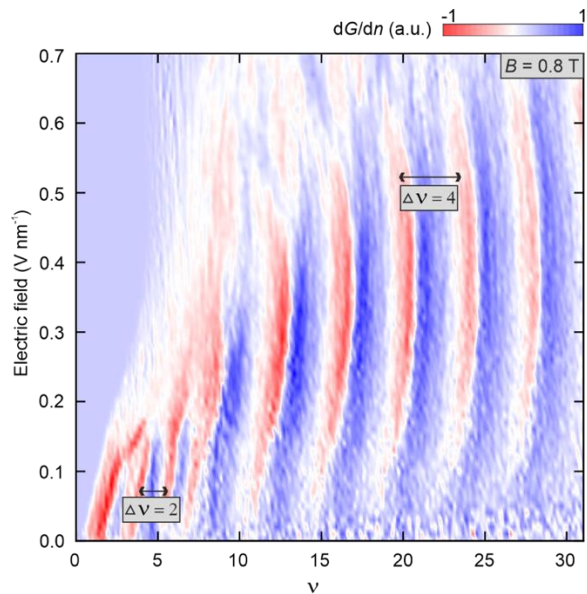

**Supplementary Fig. 2. Stoner physics in the conduction band.** Derivative of the conduction as a function of the filling factor ($\nu$) and the electric field at $B = 0.8$ T for positive filling factors. Two-fold and four-fold Landau level degeneracies are marked.

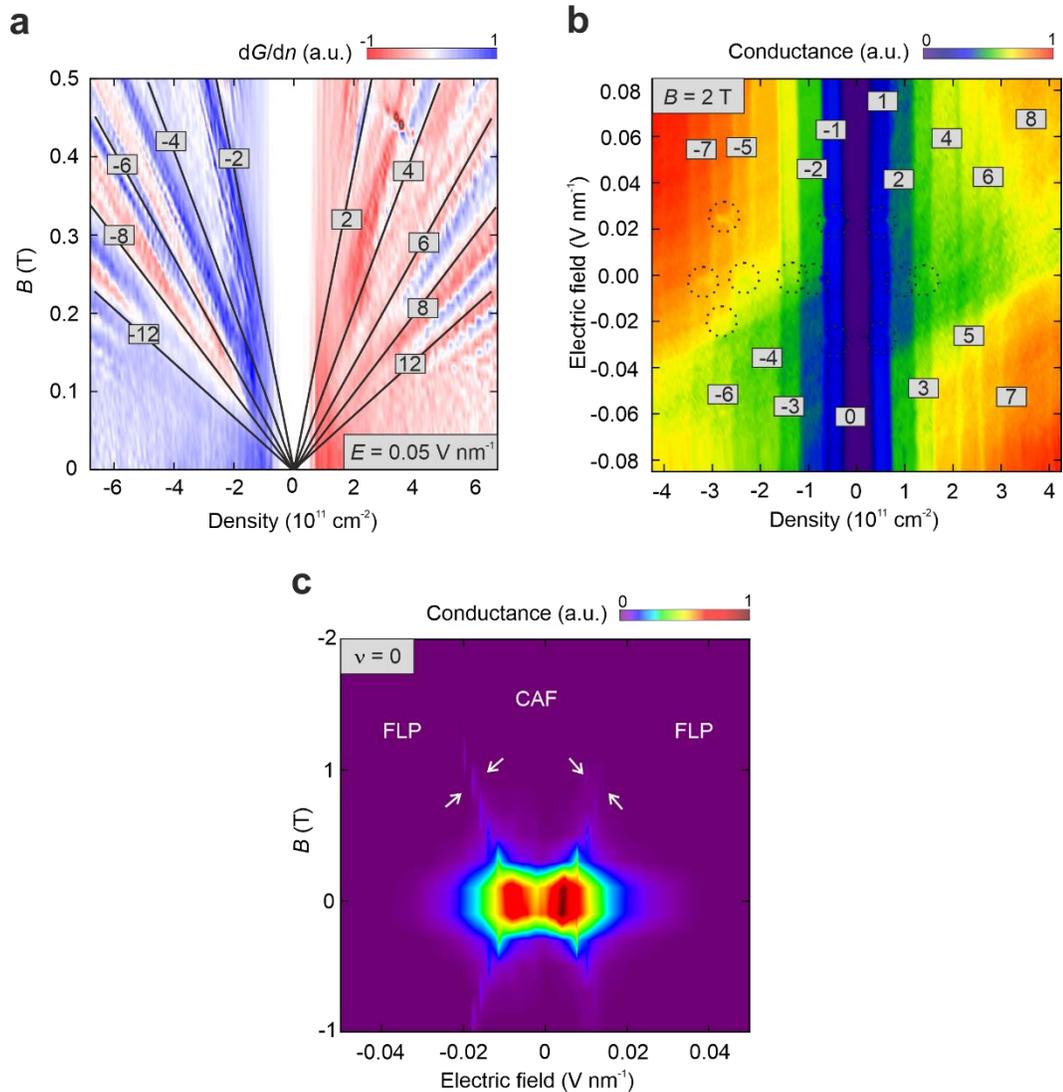

**Supplementary Fig. 3. Magnetotransport at low electric fields. (a)** Fan Diagram at $E$ = 0.05 V/nm. The slopes of the integer QH states are traced by lines and labeled by numerals. **(b)** Conductance as a function of the charge carrier density and the electric field at $B$ = 2 T. Integer QH states are labeled by numerals. Transitions induced by the electric field are marked by dashed circles. **(c)** Conductance as a function of the electric field and the magnetic field at $\nu$ = 0. The phase transitions between canted antiferromagnetic (CAF) and fully layer polarized (FLP) phases are indicated by arrows.

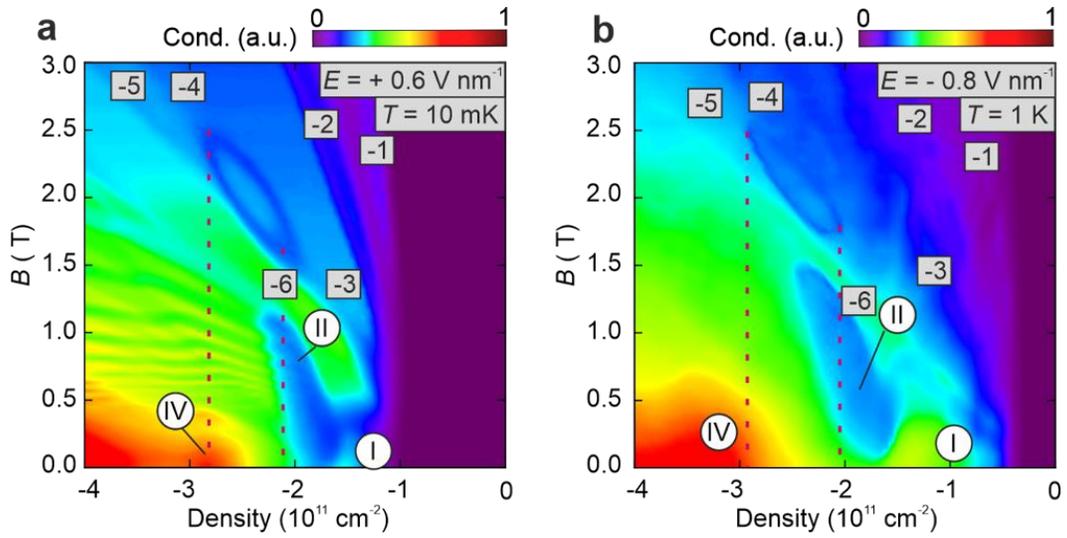

**Supplementary Fig. 4. Crossings between the $\nu$ = -4 and -5 quantum Hall states.** Fan diagrams of the conductance with respect to the charge carrier density at $E$ = 0.6 V/nm and $T$ = 10 mK **(a)** and $E$ = -0.8 V/nm and $T$ = 1 K **(b)**. Integer QH states are labeled by arabic numerals. Different phases are labeled by roman numerals. The corresponding densities of the two crossings between the $\nu$ = -4 and -5 QH states are traced to zero magnetic field by dashed lines.

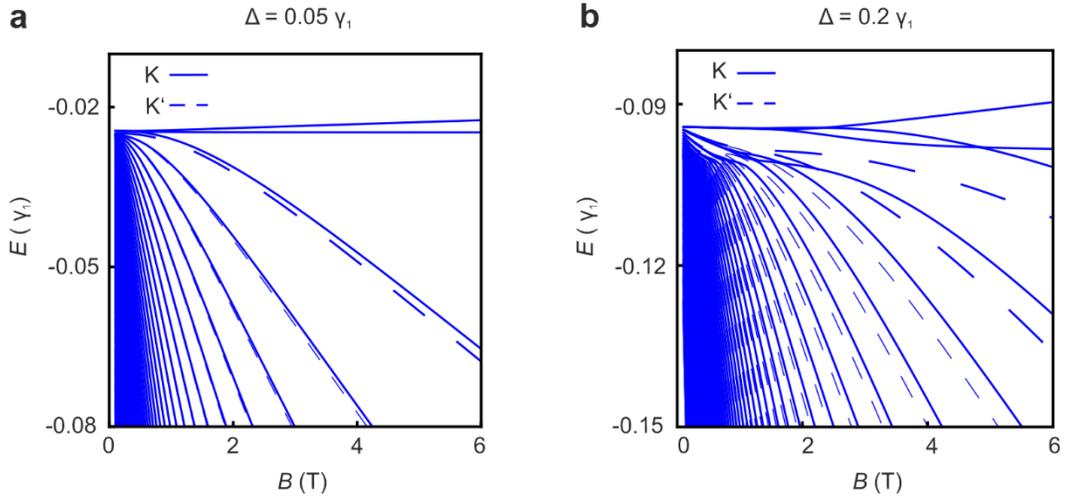

**Supplementary Fig. 5. Calculated Landau level spectra for $\Delta = 0.05\gamma_1$ (a) and $\Delta = 0.2\gamma_1$ (b).** Solid (dashed) lines correspond to the case for valley K (K'). Δ is the screened interlayer potential difference induced by an applied perpendicular electric field. $\gamma_1 \approx 400$ meV is the interlayer nearest-neighbor coupling.

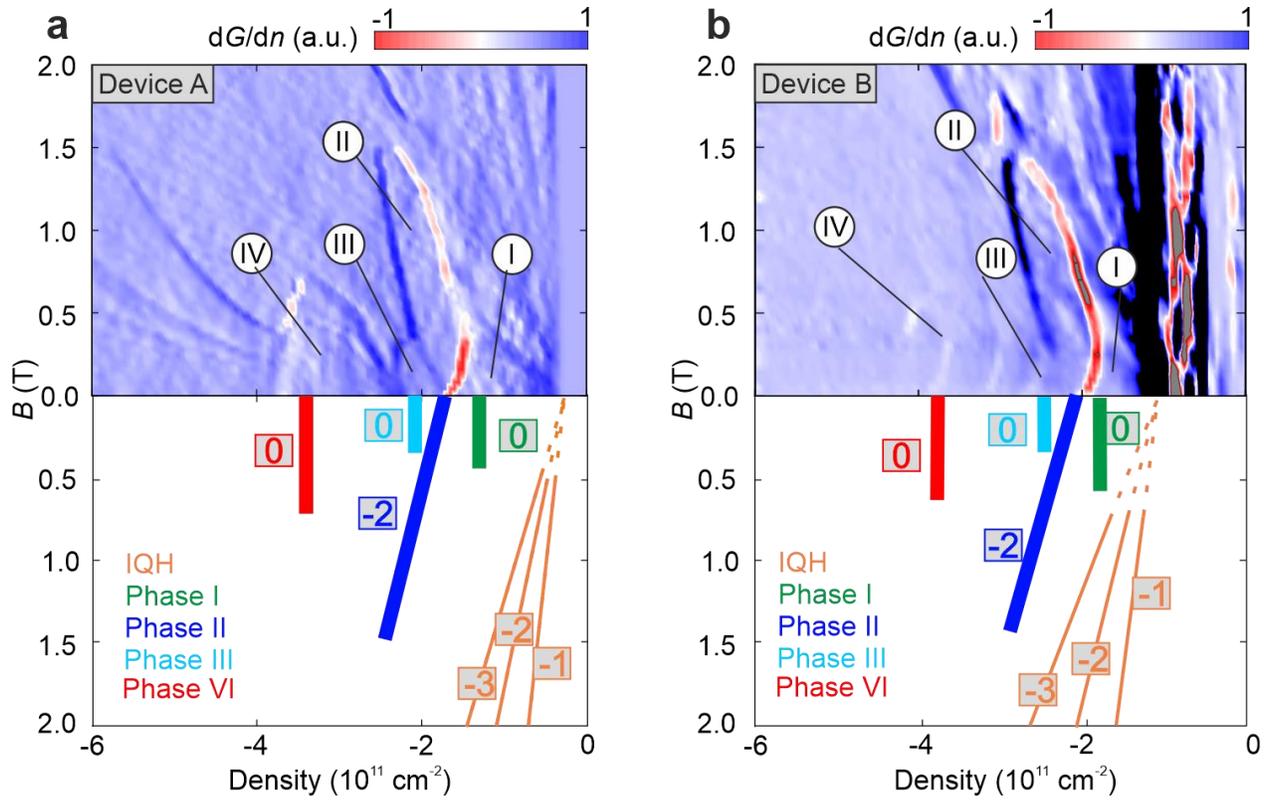

**Supplementary Fig. 6. Magnetotransport data of a second device.** Derivative of the conductance plotted as a function of the charge carrier density and the magnetic field at $E = -0.8$ V/nm for Device A **(a)** and Device B **(b)**. The slopes of the lowest integer QH states and of the phases I - IV discussed in the main text are traced by lines in the mirror images of the fan diagrams. The corresponding slopes are indicated by arabic numerals. The lines are solid if the states are present and dashed otherwise.

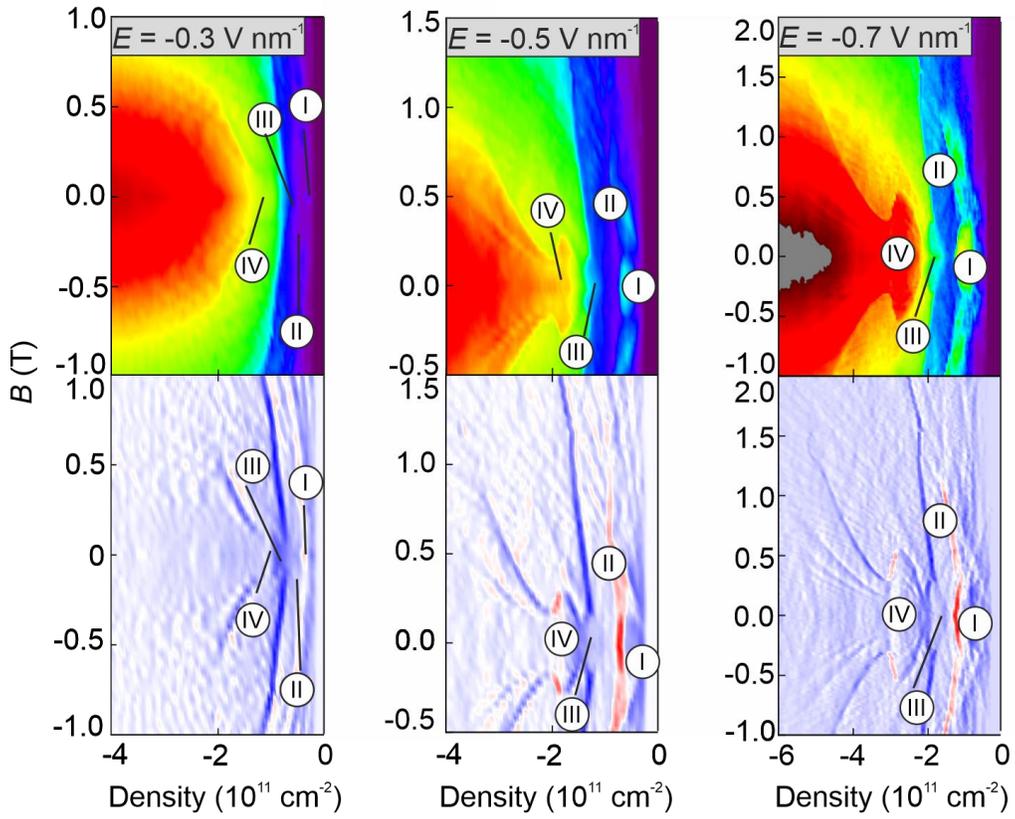
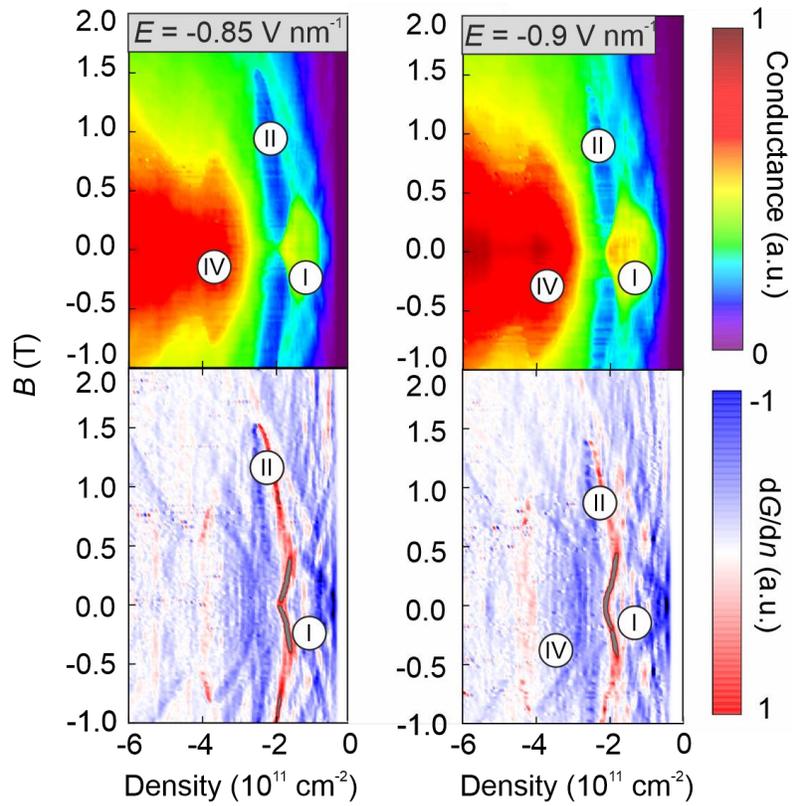

**Supplementary Fig. 7. Additional magnetotransport data.** Conductance and derivative of conductance plotted as functions of the charge carrier density and the magnetic field at different electric fields.

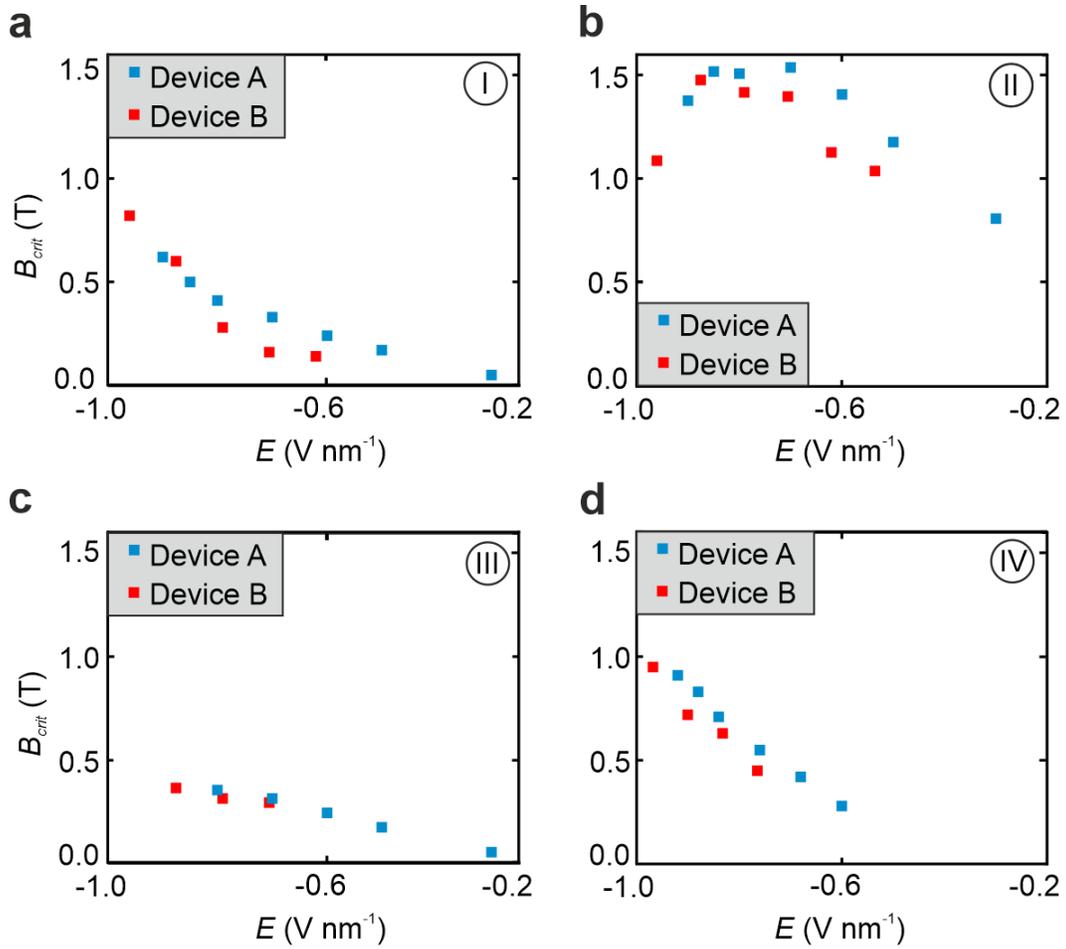

**Supplementary Fig. 8. Critical magnetic fields.** Critical magnetic fields for devices A and B of phase I **(a)**, phase II **(b)**, phase III **(c)**, and phase IV **(d)** at different electric fields.

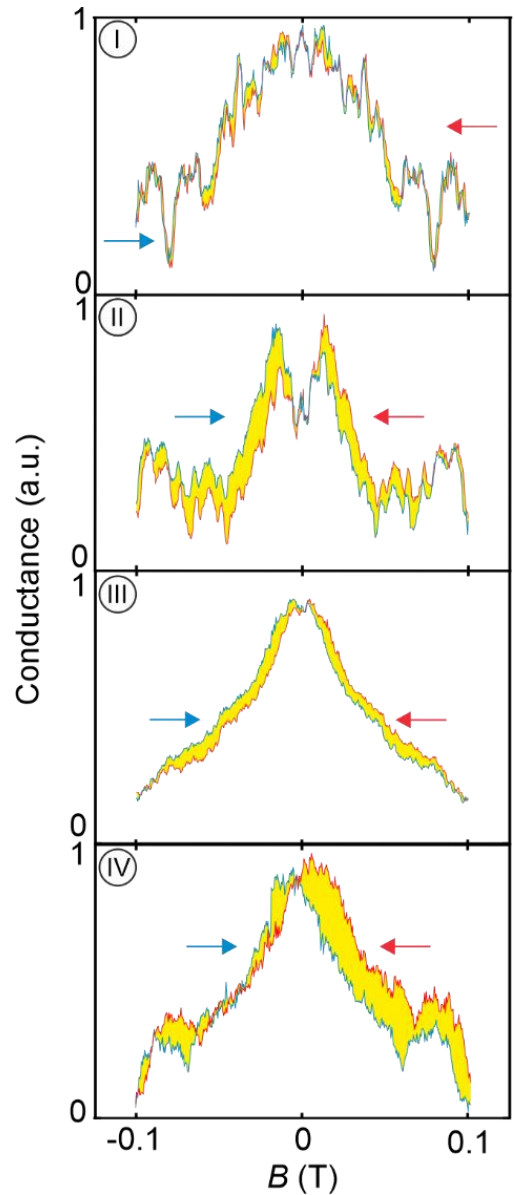

**Supplementary Fig. 9. Hystereses of new correlated phases I - IV.** Hystereses of the conductance as a function of the magnetic field at $E$= -0.6 V/nm and a charge carrier density corresponding to phase I ($n$ = -0.85 x $10^{11}$ cm$^{-2}$), phase II ($n$ = -1.2 x $10^{11}$ cm$^{-2}$), phase III ($n$ = -1.5 x $10^{11}$ cm$^{-2}$), and phase IV ($n$ = -2.2 x $10^{11}$ cm$^{-2}$), respectively. The forward sweeps are shown in blue while the reverse ones in red. The hysteresis loop areas are shaded in yellow for clarity.

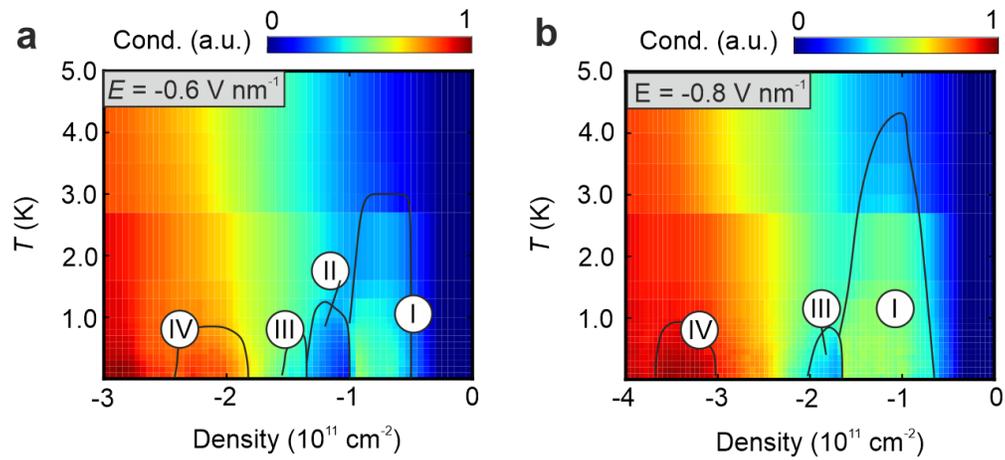

**Supplementary Fig. 10. Temperature dependence of new correlated phases I - IV.** Conductance as a function of charge carrier density and temperature at $B = 0$ and $E = -0.6$ V/nm **(a)** and $B = 0$ and $E = -0.8$ V/nm **(b)** in a larger temperature range than displayed in Fig. 4.

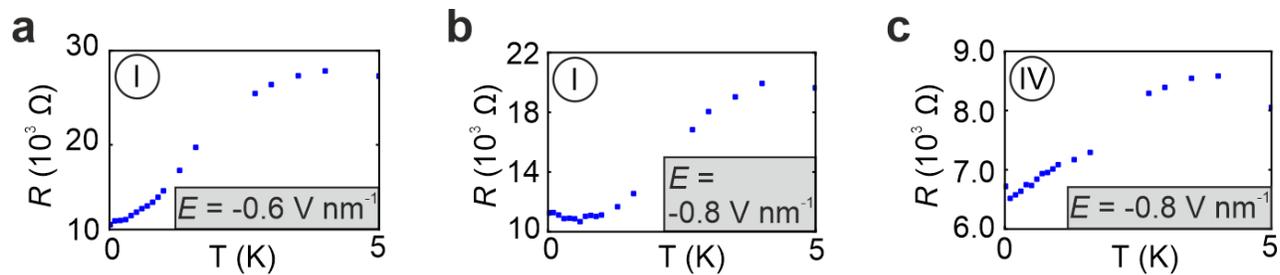

**Supplementary Figure 11: Details of the temperature dependence of phases I and IV.** The temperature evolution of the resistance shows neither quadratic nor linear.

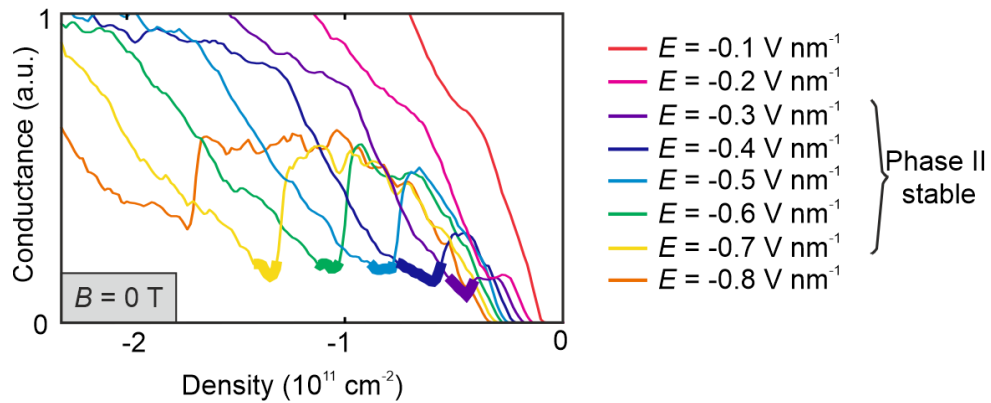

**Supplementary Fig. 12. Conductance of phase II at various different electric fields.**

Conductance as a function of charge carrier density at different electric fields and $B = 0$. Density regions of stable phase II are highlighted.